\documentclass[
superscriptaddress,
twocolumn,
bibnotes,
amsmath,amssymb,
aps,
prb,
]{revtex4-2}

\usepackage{graphicx}
\usepackage{dcolumn}
\usepackage{bm}

\usepackage{hyperref}
\usepackage{subfigure}
\usepackage{siunitx}
\usepackage{xcolor}

\begin{document}

\preprint{}

\title{Chiral-induced angular momentum radiation in single molecular junctions}

\author{Bing-Zhong Hu}
 \affiliation{School of Physics, Institute for Quantum Science and Engineering and Wuhan National High Magnetic Field Center,\\ Huazhong University of Science and Technology, Wuhan 430074, People's Republic of China}
\author{Zu-Quan Zhang}
    \affiliation{Department of Physics, Zhejiang Normal University, Jinhua 321004, People's Republic of China}
\author{Lei-Lei Nian}
    \affiliation{School of Physics and Astronomy, Yunnan University, Kunming 650091, People's Republic of China}
\author{Jing-Tao L\"u}
    \email{jtlu@hust.edu.cn}
    \affiliation{School of Physics, Institute for Quantum Science and Engineering and Wuhan National High Magnetic Field Center,\\ Huazhong University of Science and Technology, Wuhan 430074, People's Republic of China}

\date{\today}
\begin{abstract}
We study angular momentum radiation from electrically-biased chiral single molecular junctions using the nonequilibrium Green's function method. Using single helical chains as examples, we make connections between the ability of a chiral molecule to emit photons with angular momentum to the geometrical factors of the molecule. We point out that the mechanism studied here does not involve the magnetic moment. Rather, it relies on inelastic transitions between scattering states originated from two electrodes with different chiral properties and chemical potentials. The required time-reversal symmetry breaking is provided by nonequilibrium electron transport. 
Our work sheds light on the relationship between geometrical and optoelectrical chiral properties at single molecular limit. 
\end{abstract}

\maketitle

\section{\label{sec:Intro} Introduction}
Angular momentum(AM) is a fundamental property of light\cite{poynting_wave_1909,beth_mechanical_1936,coullet_optical_1989,allen_orbital_1992,bliokh_transverse_2015,shen_optical_2019}, whose generation and manipulation are of vital importance for their applications in optoelectronics, quantum information science, and so on\cite{jack_holographic_2009,bozinovic_terabit-scale_2013,andrews_optical_2004,krenn_generation_2014,shen_optical_2019}.
Light with AM can be generated by physical objects with vastly different scales, from as small as synchrotron in particle physics\cite{katoh_helical_2017,katoh_angular_2017,epp_angular_2019,lan_electron_2020} to as large as rotating black hole in astrophysics\cite{tamburini_twisting_2011}. The AM of light can be furthermore used to study other types of chiral excitations.

The magneto-electric coupling, depending on both the magnetic and electric dipole transition matrix elements, is a key factor that determines the magnitude or efficiency of many of the above-mentioned processes\cite{polavarapu_chiroptical_2017}. Unfortunately, the magnetic dipole transition is much weaker than the corresponding electric one, resulting in a small magneto-electric coupling. Employing the chiral geometric or electronic structure in electric dipole transitions, akin to the so-called extrinsic chirality, is a promising approach to avoid the weak magnetic dipole transition.

Recently, it has been shown theoretically that coupling of electron orbital motion with light in current-carrying molecular junctions can lead to AM radiation (AMR)\cite{zhang_angular_2020,zhang_far-field_2020,zhang_electroluminescence_2021,ridley_quantum_2021,zhang_microscopic_2022}. Electroluminescence from single molecules or localized gap plasmons in a scanning tunneling microscope (STM) has been studied for decades \cite{berndt1993photon,qiu2003vibrationally,dong_generation_2010,kazuma_real-space_2018,doppagne_electrofluorochromism_2018,zhang2017sub,imada2016real}. Different quantum statistical properties of emitted light has been characterized using STM setup\cite{zhang2017electrically,merino2018bimodal,leon2019photon,nian_fano_2018,nian_plasmon_2022}. Thus, it is also an ideal experimental candidate to study AMR at the single molecular scale.

On the other hand, molecular electronics and optoelectronics using chiral molecules have been the focus of recent intense research\cite{zhang_recent_2020}. In the phenomenon of chiral-induced spin selectivity (CISS)\cite{guo_spin-selective_2012,guo_spin-dependent_2014,gutierrez2012spin,gutierrez2013modeling,gohler_spin_2011,naaman_chiral-induced_2012,dalum_theory_2019,naaman_chiral_2020,liu_chirality-driven_2021,evers_theory_2022,naskar_common_2022}, spin-polarized electrons can be generated from chiral molecular structure driven by electrical or optical stimuli. Spin-orbit interaction is argued to play an important role, although the exact mechanism is still under debate. In light emitting diode, large chiroptical effects are observed from chiral molecular structures\cite{greenfield_pathways_2021,liu_chirality-driven_2021}. Its origin is attributed to either the magneto-electric coupling (natural optical activity), or structural chirality. Notably, a recent work proposed an electronic mechanism employing the topological electronic structure for circular polarized light emission under electrical  current flow\cite{wan_anomalous_2023}. 
The common trends of CISS and optical dichroism in helical structures is also studied very recently\cite{naskar_common_2022}.

In this work, we study theoretically AMR from junctions of model helical chains using nonequilibrium Green's function (NEGF) method\cite{haug2008quantum,zhang_angular_2020}. We analyze how the chiral molecular geometry can directly lead to AMR. We trace its origin as inelastic scattering between scattering states originated from different electrodes, which does not rely on the spin-orbit coupling. Furthermore, we study the effect of length, radius and other parameters of the chain on the emission spectrum. 
Suitable conditions for enhancing the AMR are proposed based on the numerical calculation. This study will be useful for the design of chiral light sources based on single molecular junctions.

\begin{figure}[b]
\includegraphics[width=\linewidth]{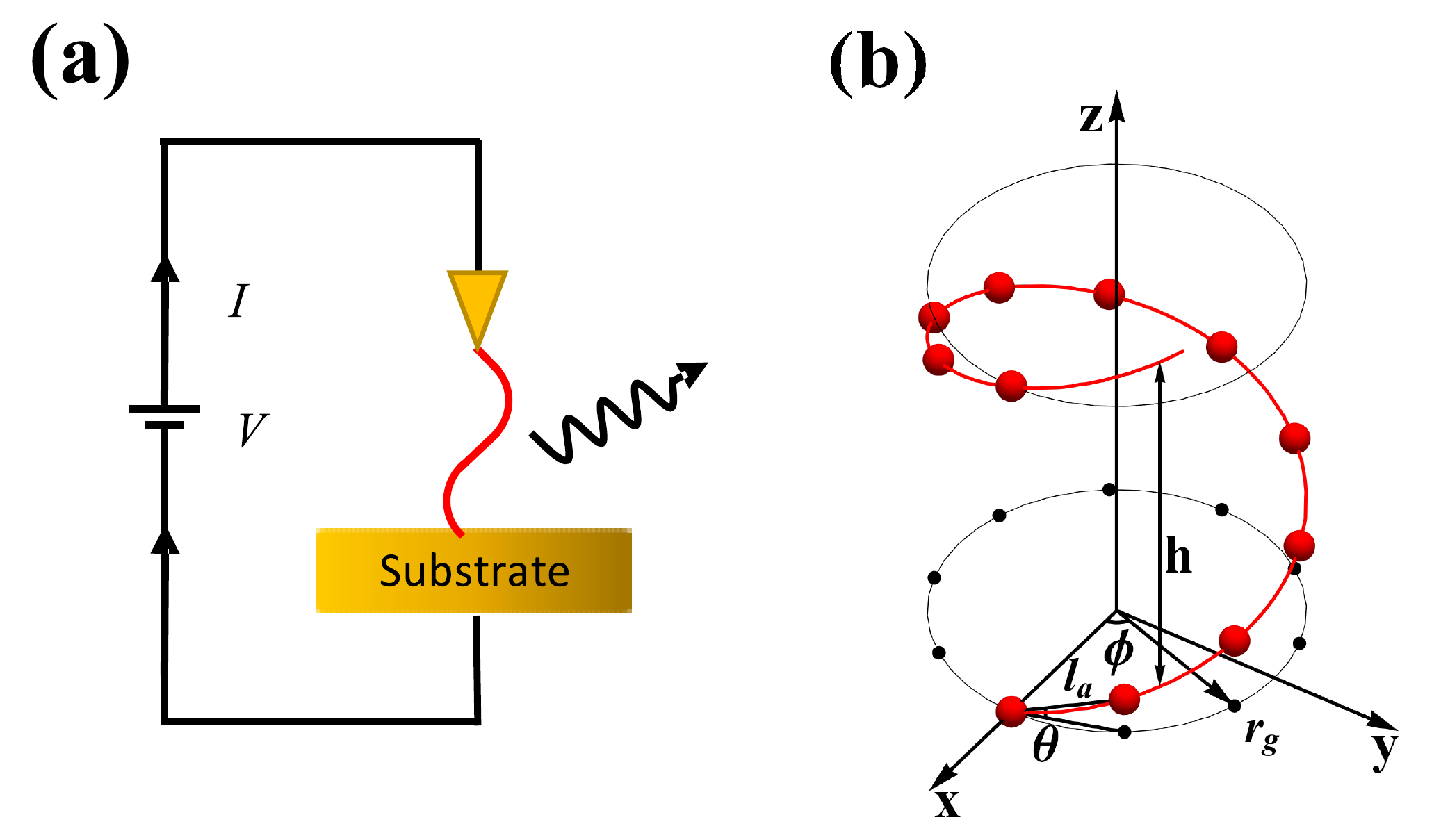}

\caption{(a) Schematic diagram of angular momentum radiation (AMR) from voltage-biased chiral molecule. (b) A helical chain structure with radius of gyration $r_g$. The pitch of a helix $h$ describes the vertical distance raised in $z$ direction after $2\pi$ rotation in $\phi$, while $\theta$ represents angle between the line connecting two neighbouring sites and the $x$-$y$ plane. Neighboring sites are separated by a distance of $l_a$. 
}
\label{fig:epsart}
\end{figure}

\section{\label{sec:thoery}Model and theory}
\subsection{Hamiltonian}
We use a tight binding model to write the Hamiltonian of the molecule as \cite{zhang_angular_2020,zhang_far-field_2020}
\begin{align}
    H_{\rm mol} = \sum_{mn} H_{mn}c_m^\dagger c_n e^{i\theta_{mn}} ,
\end{align}
where $c_m^\dagger/c_m$ is the creation/annihilation operator at site $m$, $H_{mn}$ is the hopping matrix element between sites $m$ and $n$.
Molecular coupling to the radiation fields is taken into account by the Peierls substitution\cite{graf_electromagnetic_1995} with the phase factor $\theta_{mn}$ given by an integral along the line connecting sites $m$ and $n$
\begin{align}
\theta_{mn}=\frac{e}{\hbar} \int_{\vec{r}_n}^{\vec{r}_m} \vec{A} \cdot d \vec{l} \approx \frac{e}{2\hbar}(\vec{A}_m+\vec{A}_n)\cdot(\vec{r}_m-\vec{r}_n).
\end{align}
Here, we have adopted the trapezoidal rule for the integral, 
$e$ is the elementary charge, $\hbar$ is the reduced Planck constant, $\vec{r}_m$ is the position vector of site $m$, $\vec{A}_m = \vec{A}(\vec{r}_m)$ represents vector potential of the electromagnetic field at site $m$.
Hamiltonian of the radiation field is
\begin{align}
    H_{\mathrm{rad}}=\frac{1}{2} \int d^3 \vec{r}\left(\varepsilon_0 \vec{E}_{\perp}^2+\frac{1}{\mu_0} \vec{B}^2\right),
\end{align}
where $\varepsilon_0, \mu_0$ are the vacuum permittivity and permeability, respectively.  $\vec{E}_\perp$ and $\vec{B}$ are electric field and magnetic field in the transverse gauge $\nabla \cdot \vec{A}=0$. They are written in terms of $\vec{A}$ as  $\vec{E}_{\perp}=-\partial_t \vec{A}$, $\vec{B}=\nabla \times \vec{A}$. 

By using smallness of the lattice spacing $\vec{r}_m-\vec{r}_n$, we can expand the exponential form of Peierls substitution to the first order in $\vec{A}$ and divide the molecule Hamiltonian into two terms
\[ H_{\rm mol} = H_0 + H_{\rm int}. \]
The first term $H_0$ is the non-interacting part
\begin{align}
    H_0 = \sum_{m,n}H_{mn}c_m^\dagger c_n,
\end{align}
and the second term $H_{\rm int}$ represents coupling of electrons with electromagnetic field represented by the vector potential 
\[H_{\rm int} \approx \sum_{mn} \sum_k \sum_{\mu=x, y, z} M_{mn}^{k \mu} c_m^{\dagger} c_n A_\mu\left(\vec{r}_k\right), \]
with
\[M_{mn}^{k \mu}= \frac{i e}{2 \hbar} H_{mn}\left(\vec{r}_m-\vec{r}_n\right)_\mu\left(\delta_{k m}+\delta_{k n}\right)\]
the electron-photon coupling matrix element. 

Since the main purpose of this work is to explore the physical mechanism of AMR, for the sake of simplicity, we make the following two simplifications. First, we ignore the complicated spatial dependence of the vector potential due to presence of molecular junction and use the distribution in vaccum. By doing this, we have ignored the local enhancement of the electromagnetic field due to plasmon modes. 
Second, we ignore spin degrees of freedom and spin-orbit coupling to highlight that the physical origin of AMR is the orbital motion of electrons, and does not rely on the spin degree of freedom of electrons. Mutual coupling between electron spin, orbital motion and light polarization may lead to new physical effect, but it is out of the scope of present study. 
Third, we take the electronic subsystem as a source of the electromagnetic radiation and ignore the back-action of the electromagnetic field on electrons. In this way, we work on the lowest order level in dealing with the photon self-energy due to coupling to electrons.

\subsection{Electron and photon Green's functions}
In this subsection, we give the definition of the electron and photon Green's functions, which will be used  in next subsection to calculate the energy and angular momentum current. Details of the nonequilibrium Green's function method can be found in several textbooks and review articles\cite{haug2008quantum,wang_quantum_2008-1,wang2023transport}. 

The electron and photon contour time ordered Green's functions are defined as
\begin{align}
    G_{mn}(\tau,\tau') &= \frac{1}{i\hbar}\langle T_{\tau} c_m(\tau) c_n^\dagger (\tau') \rangle , \label{eq:defg}\\
    D_{\mu\nu}(\vec{r}_m,\tau;\vec{r}_n,\tau') &= \frac{1}{i\hbar}\langle T_{\tau} A_\mu(\vec{r}_m,\tau) A_\nu^\dagger (\vec{r}_n,\tau') \rangle .  \label{eq:defd}
\end{align}
The indices in the electron Green's function $G_{mn}$ are located in the molecule. Meanwhile, the coordinates in the photon Green's functions can be in any point in real space $\vec{r}$, with $\mu$ and $\nu$ taking $x$, $y$ or $z$. Going from the contour time ($\tau, \tau'$) to real time ($t, t'$), different types of Green's function can be generated. The retarded Green's functions are defined as
\begin{align}
    G^{r}_{mn}(t,t') & = \frac{1}{i\hbar} \Theta(t-t') \langle [c_m(t), c_n^\dagger (t')] _+ \rangle  , \label{eq:defgr}\\
    D^{r}_{\mu\nu}(\vec{r}_m,t;\vec{r}_n,t') &= \frac{1}{i\hbar} \Theta(t-t') \langle [ A_\mu(\vec{r}_m,t), A_\nu^\dagger (\vec{r}_n,t') ]_- \rangle ,  \label{eq:defdr}
\end{align}
with $[A_\mu,A^\dagger_\nu]_{\pm} = A_\mu A^\dagger_\nu \pm A^\dagger_\nu A_\mu$ and $\Theta(t)$ the Heaviside step function. In the nonequilibrium case, we will also need the lesser Green's functions
\begin{align}
G^{<}_{mn}(t,t') & = -\frac{1}{i\hbar} \langle c_n^\dagger (t') c_m(t)\rangle, \label{eq:defgl} \\
D^{<}_{\mu\nu}(\vec{r}_m,t;\vec{r}_n,t') &= \frac{1}{i\hbar} \langle A_\nu^\dagger (\vec{r}_n,t') A_\mu(\vec{r}_m,t)  \rangle .  \label{eq:defdl}
\end{align}
At steady state considered in this work, the Green's functions only depend on the time difference $t-t'$ and they can be transformed to frequency/energy space via Fourier transform, i.e., 
\begin{align}
    G_{mn}^r(E) = \int dt~G_{mn}^r(t-t') e^{iE (t-t')/\hbar}.
\end{align}
Effect of the coupling to the electrodes can be taken into account via self-energies.  The electron lesser and greater Green's functions are then obtained as
\begin{align}
    G^{</>}(E) = G^r(E) \Sigma^{</>}(E) G^a(E),
\end{align}
with embedding electron self-energy $\Sigma = \Sigma_L + \Sigma_R$. Define the electrode broadening function ($\alpha=L, R$) 
\begin{align}
    \Gamma_\alpha (E) &= i(\Sigma_\alpha^r (E) - \Sigma_\alpha^a (E)), \\
    \Gamma (E) &= \Gamma_L (E)+\Gamma_R (E),\\
    A_\alpha (E) &= G^r (E) \Gamma_\alpha (E) G^a (E), \\
    A (E)&= A_L (E)+A_R (E),
\end{align}
they can be further written as
\begin{align}
    G^{<} (E) &= -\sum_\alpha f_\alpha (E) (G^r (E)-G^a (E)) \nonumber\\
    &= i\sum_\alpha f_\alpha (E) A_\alpha (E) , \label{eq:ggla1} \\
    G^{>} (E) &= \sum_\alpha (1-f_\alpha (E)) (G^r (E)-G^a (E)) \nonumber\\
    &= -i\sum_\alpha (1-f_\alpha (E)) A_\alpha (E) . \label{eq:ggla2}
\end{align}
Here $f_\alpha (E)$ is the Fermi-Dirac distribution function
\begin{align}
f_\alpha(E)=1/[{\rm exp}((E-\mu_\alpha)/k_BT)+1] .
\end{align}

\begin{figure}[ht]
    \centering
    \includegraphics[width=0.4\linewidth]{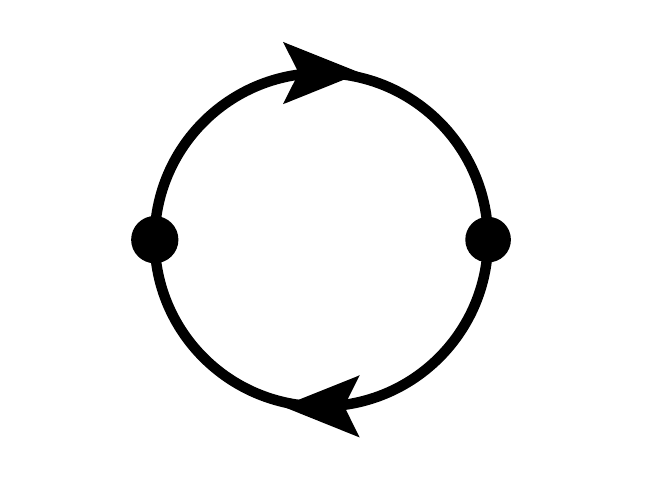}

    \caption{Feynmann diagram of photon self energy, where the solid lines with arrows represent electrons.} 
    \label{fig:photon_diagram}
\end{figure}

To consider electrically driven energy and angular momentum radiation, we take into account electron-photon coupling in the lowest order. 
The retarded GF of photons is given by the Dyson equation 
\begin{align}
D^{r}(\omega) = d^{r}(\omega) + d^{r}(\omega) \Pi^{r}(\omega) D^{r}(\omega)
\label{eq:dyson1}
\end{align}
with $d^{r}(\omega)$ the free space photon GF, $\Pi^{r}(\omega)$ photon retarded self-energy due to coupling to electrons in the random phase approximation (RPA). The lesser GF is calculated from the equation 
\begin{align}
D^{<}(\omega) = D^{r}(\omega) \Pi^{<}(\omega) D^{a}(\omega)
\label{eq:keldysh}
\end{align}
with $D^{a} = (D^{r})^{\dag}$. To the lowest order in electron-photon coupling, we consider only the photon self-energy shown in Fig. \ref{fig:photon_diagram}, written as
\begin{align}
    \Pi^{<}_{\mu\nu}(\vec{r}_m,\vec{r}_n,\omega) =-i \int_{-\infty}^{\infty} \frac{d E}{2 \pi} \operatorname{Tr}\left[M^{m \mu} G^{<}(E) M^{n \nu} G^{>}(E^-)\right].
\end{align} 
Here, ${\rm Tr}[...]$ is trace over all electronic degrees of freedom, and $E^-=E-\hbar \omega$. Although the photon Green's function is defined in the whole space, the above self-energy is defined only on the discrete sites of the molecule. 
In principle, the electron and photon Green's functions and self-energies should be calculated self-consistently. Here, we avoid this complication and calculate the photon self-energy in the lowest order, using the noninteracting version of $G^{</>}$.
In doing so, as we have mentioned, we can not take into account the collective excitation of electrons, i.e., plasmons. In reality, localized plasmon modes may play important roles in enhancing the light emission yields in single molecular junctions. 

\subsection{Electronic transport }
In the noninteracting limit, the electrical current can be obtained from the Landauer formula
\begin{align}
    I = \frac{e}{h}\int dE \mathcal{T}(E) (f_L(E) - f_R(E)) .
\end{align}
with the electron transmission function
\begin{align}
    \mathcal{T}(E) = {\rm Tr}[G^r(E)\Gamma_L(E) G^a(E) \Gamma_R(E)] .
    \label{eq:trans}
\end{align}

\subsection{Light and angular momentum radiation}
We focus on the electromagnetic radiation at the far field, and closely follow the method presented in Ref.~\onlinecite{wang2023transport}, to which we refer for the details. 
The total energy $U$ and angular momentum $\vec{L}$ of an electromagnetic radiation field (light) are defined as 
\begin{eqnarray}
 U &= &\frac12 \int (\varepsilon_0 \vec{E}^2 + \frac{1}{\mu_0}\vec{B}^2) d^3 \vec{r}, \\
\vec{L} &= &\frac{1}{c^2} \int \vec{r} \times \vec{S} d^3 \vec{r},
\end{eqnarray}
with $c=(\varepsilon_0\mu_0)^{-1/2}$ the speed of light in vacuum, and 
the Poynting vector 
\begin{align}
\vec{S}=\frac{1}{\mu_0}\vec{E}_\perp \times \vec{B}.
\end{align}
From the conservation laws of energy and angular momentum,
the energy or angular momentum current can be obtained 
\begin{eqnarray}
P &=  \oint \vec{S} {\cdot} d\vec{\Omega}, \label{eq:RadPower} \\
\vec{J} &=  \oint \overset\leftrightarrow{\mathcal{M}} {\cdot} d\vec{\Omega},  \label{eq:AMFPower}
\end{eqnarray}
with the surface vector $\vec{\Omega}$, 
the AM flux
\begin{eqnarray}
\overset\leftrightarrow{{\mathcal{M}}} &=& \vec{r} \times \overset\leftrightarrow{{T}}, \label{eq:AMF}
\end{eqnarray}
where
$\overset\leftrightarrow{{T}}$ is the Maxwell stress tensor  with 
\begin{align}
T_{\mu \nu} = \frac{1}{2} \delta_{\mu \nu} (\varepsilon_{0} E_\perp^2 + \mu_{0}^{-1} B^2) - \varepsilon_0 E_{\perp\mu} E_{\perp\nu} - \mu_{0}^{-1} B_{\mu} B_{\nu} .
\end{align}
These quantities can be quantized and expressed in terms of the photon GFs as follows

\begin{subequations}
\begin{align}  
\begin{split}
S^{\mu} (\vec{r})  =&\epsilon_{\mu \nu \gamma} \epsilon_{\gamma \delta \xi}  \frac{2}{\mu_{0}}  \int_{0}^{+\infty} \frac{d \omega}{2 \pi} \hbar \omega     \\
 \times &\textrm{Re}\bigg[- \frac{\partial}{\partial x_{\delta}' } D_{\nu \xi}^{<}(\vec{r},\vec{r}' , \omega)  \bigg] \bigg{|}_{\vec{r}' \to \vec{r}}, \label{eq:Sperp}
\end{split}  \\
 \langle \colon E_{\perp\mu} E_{\perp\nu} \colon \rangle = & \textrm{Re} \big[ \frac{2 i}{\hbar} \int_{0}^{\infty} \frac{d\omega}{2 \pi} (\hbar \omega)^2  
 \times    D_{\mu \nu}^{<}(\vec{r},  \vec{r} , \omega) \big],  \label{eq:EEGF}   \\
\begin{split}
    \langle \colon B_{\mu} B_{\nu} \colon \rangle = &  \textrm{Re} \bigg[ 2i  \hbar \int_{0}^{\infty} \frac{d \omega}{2 \pi} \epsilon_{\mu \gamma \xi}  \epsilon_{\nu \gamma' \xi'} \\
    \times &  \frac{\partial}{\partial x_{\gamma}}  \frac{\partial}{\partial x_{\gamma'}' }  D_{\xi \xi'}^{<}(\vec{r},  \vec{r}' , \omega) \bigg] \bigg{|}_{\vec{r}' \to \vec{r}}. \label{eq:BBGF}
\end{split}
\end{align}
\end{subequations}
Here, $\langle \colon AB \colon \rangle$ denotes normal order of operators $A$ and $B$ when taking the ensemble average. The purpose is to remove the zero point motion contribution\cite{wang2023transport}. 
The Greek subscript letters $\mu$, $\nu$, $\gamma$ are indices for the Cartesian coordinates, and 
$\epsilon_{\mu \nu \gamma}$ is the Levi-Civita symbol 
with $\epsilon_{xyz}=1$, antisymmetric with each permutation of any two indices, and zero if any two of the indices are the same.

Approximations need to be made to get simplified results. First, a lowest order approximation is taken via replacing the photon's retarded and advanced GFs by the free space photon GFs with $D^r\approx d^r$ and  $D^{<} \approx d^{r} \Pi^{<} d^{a}$. This means we take the electromagnetic field as unperturbed and ignore the local modification near the single molecule due to presence of molecular junction. The effect of localized plasmons is thus ignored. It can in principle be included by using the full equations [Eqs.~(\ref{eq:dyson1},\ref{eq:keldysh})], but is beyond the scope of current work. Second, a monopole approximation\cite{wang2023transport} is taken when calculating the surface integral in the far-field limit $\vec{r} \rightarrow \infty$, by using $d_{\mu \nu}^{r} (\vec{r}, \vec{r}_{m}, \omega) = d_{\mu \nu}^{r} (\vec{r} - \vec{r}_{m}, \omega)   \approx d_{\mu \nu}^{r} (\vec{r}, \omega)$ for the free space GF, which reads
\begin{align}
    \overset\leftrightarrow{d^r}(\vec{r}, \omega) = &-\frac{1}{4\pi\varepsilon_0c^2r}\left\{e^{i\frac{\omega}{c}r}(\overset\leftrightarrow{U}-\hat{R}\hat{R})\right.\\
    &\left.+\left[-\frac{e^{i\frac{\omega}{c}r}}{i\frac{\omega}{c}r}+\frac{e^{i\frac{\omega}{c}r}-1}{(i\frac{\omega}{c}r)^2}\right](\overset\leftrightarrow{{U}}-3\hat{R}\hat{R})\right\},\nonumber
\end{align}
with $\overset\leftrightarrow{U}$ the identity dyadic, $\hat{R} = \vec{r}/r$ the radial direction unit vector.
Thus, the atomic structure of the molecule is also ignored. With these approximations, we can calculate 
the radiated power $P$, angular momentum current in $\gamma$ direction $J_\gamma$ and the photon number current $J_N$ in the far field.
This is realized by introducing `bath at infinity'. The details of this treatment can be found in Ref.~\cite{wang2023transport}. The basic idea is to include a self-energy that represents a bath at infinity. This bath locates far away from the source and can absorb all the radiated photon. The emitted power, angular momentum can then be calculated using the standard NEGF method as that injected into this bath.
The following results are then obtained \cite{zhang_angular_2020,wang2023transport}
\begin{align}
P &=-\sum_{\mu}\int_{0}^{\infty} \frac{d \omega}{2 \pi} \frac{\hbar \omega^{2}}{3 \pi \varepsilon_{0} c^{3}} \operatorname{Im}\left[\Pi_{\mu \mu}^{\mathrm{tot},<}(\omega)\right],  \label{eqn:Power}\\
J_\gamma &=\sum_{\mu,\nu}\int_{0}^{\infty} \frac{d \omega}{2 \pi} \frac{\hbar \omega}{3 \pi \varepsilon_{0} c^{3}} \epsilon_{\gamma \mu \nu} \operatorname{Re}\left[\Pi_{\mu \nu}^{\mathrm{tot},<}(\omega)\right], \\
J_N &=-\sum_{\mu}\int_{0}^{\infty} \frac{d \omega}{2 \pi} \frac{ \omega}{3 \pi \varepsilon_{0} c^{3}} \operatorname{Im}\left[\Pi_{\mu \mu}^{\mathrm{tot},<}(\omega)\right].
\label{equ:JN}
\end{align}
Here, the superscript `tot' means summation over all the sites in the system
$\Pi^{\rm tot, <}_{\mu\nu}(\omega) = \sum_{m,n}\Pi^<_{\mu\nu}(\vec{r}_m, \vec{r}_n,\omega)$.

\begin{figure*}[hbt]
\includegraphics[width=0.8\linewidth]{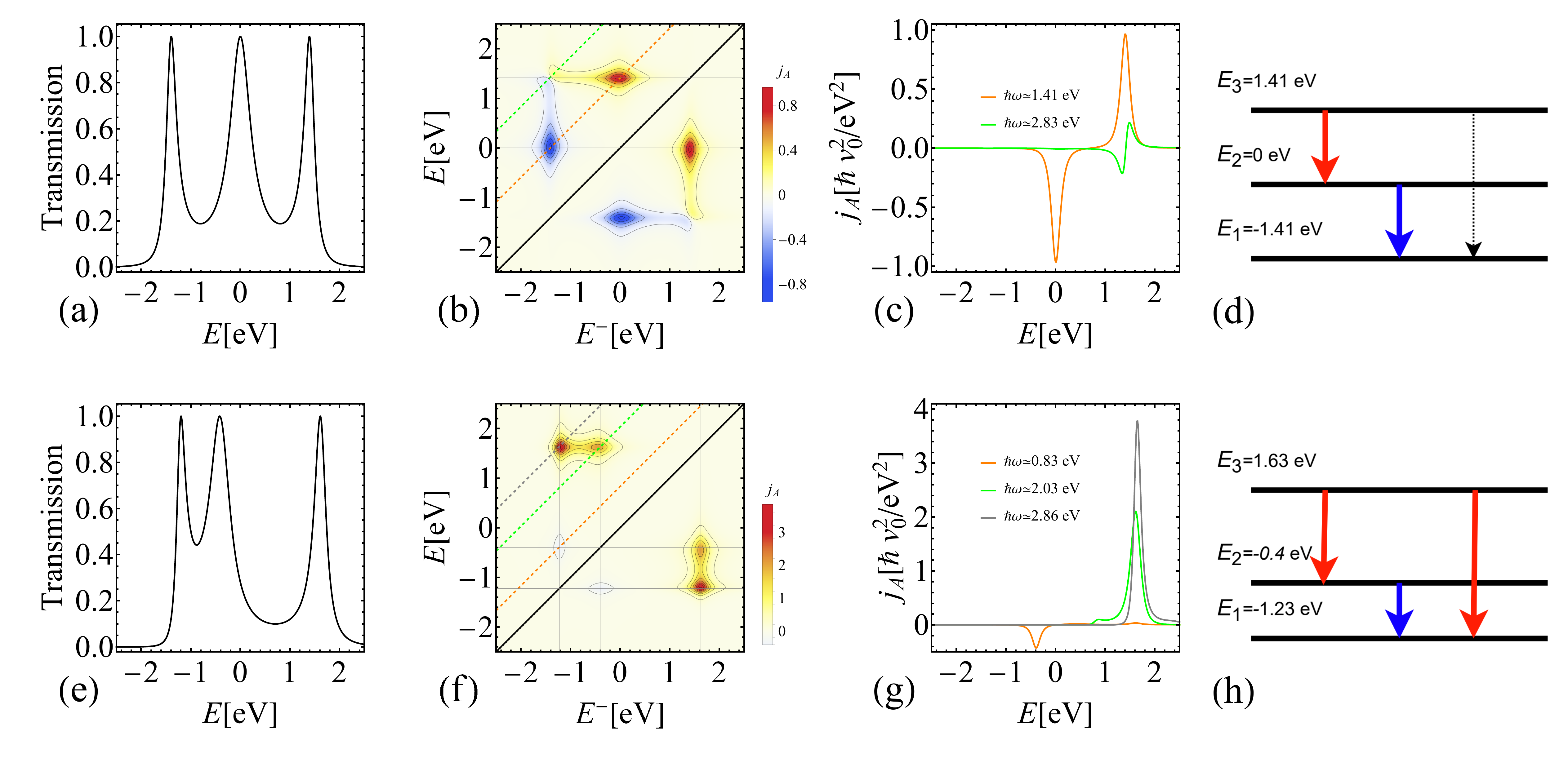}
    \caption{ Numerical results for a single helical chain with the total number of sites $N_c=3$. In the upper row, only the nearest neighbour (NN) hopping is included. In the lower row, in addition to $t_{\rm NN}$, next nearest neighbour (NNN) hopping $t_{\rm NNN}=0.4 $ eV is included. 
        (a, e) Electron transmission spectra obtained from Eq.~(\ref{eq:trans}). (b, f) $j_A$ as a function of energy $E$ and $E^-$. The velocity is $v_0 = l_a t_{\rm{NN}} /\hbar$. (c, g) Line cuts of the plot in (b, f). The corresponding energy is denoted in the insets. (d, h) Schematic diagram of inelastic electronic transitions between different molecular eigen states that may lead to light emission. Red (blue) arrow represents transition with positive (negative) AMR, while dashed arrow corresponds to zero AMR. }
    
    \label{fig:mod1}
\end{figure*}

\subsection{Physical mechanism of angular momentum radiation}
In the above subsection we have given a numerical recipe to calculate the photon and AMR using the NEGF method. We see that the key quantity is the photon self-energy due to coupling to the electronic system $\Pi^{\rm tot,<}_{\mu\nu}$, which contains all the material properties and its coupling to the electromagnetic field.  In this subsection, we reveal the physical mechanism of AMR.

To make further analysis, we use Eqs.~(\ref{eq:ggla1}-\ref{eq:ggla2}) to rewrite the self-energy as:
\begin{align}
    &\Pi^{<}_{\mu\nu}(\vec{r}_m,\vec{r}_n,\omega) =-i \sum_{\alpha,\beta}\int_{-\infty}^{\infty} \frac{d E}{2 \pi} (f_\alpha(E)-f_\beta(E^-)\nonumber\\
    &\times n_B(\hbar\omega-\Delta\mu_{\alpha\beta}) \operatorname{Tr}\left[M^{m \mu} A_\alpha(E) M^{n \nu} A_\beta(E^-)\right] .
\end{align} 
We have introduced the Bose-Einstein distribution function
\begin{align}
    n_B(\omega) = 1/[\rm{exp}(\hbar\omega/k_BT)-1] ,
\end{align}
and $\Delta \mu_{\alpha\beta}=\mu_\alpha-\mu_\beta$.
Since the photon energy is much larger than $k_BT$, we take the zero temperature limit. 
The photon lesser self-energy is simplified to 
\begin{align}\label{eq:pi2}
\Pi_{\mu \nu}^{\rm tot,<}\left( \omega\right) &= -ie^2 \sum_{\alpha,\beta=L,R}\int_{\mu_\beta+\hbar\omega}^{\mu_\alpha} \frac{dE}{2\pi}\\ &\times\Theta(\Delta\mu_{\alpha\beta}-\hbar\omega) X_{\mu\nu}^{\alpha\beta}(E,E^-). \nonumber
\end{align}
Here,  the Heaviside step function $\Theta(E)$ gives the energy range where the inelastic transitions can take place. We have defined
\begin{align}\label{eq:x2}
    &X^{\alpha \beta}_{\mu\nu}(E, E^-)=\operatorname{Tr}\left[v^\mu A^\alpha(E) v^\nu A^\beta(E^-)\right]  \nonumber\\
    &~~~~~~=(2\pi)^2 \sum_{m,n}  \delta(E-E_{m}) \delta(E^--E_{n})   \\
    &~~~~~~\times \langle \psi_{\alpha,m} (E_m) |v^\nu|\psi_{\beta,n} (E_n)\rangle\nonumber\\
    &~~~~~~\times \langle \psi_{\beta,n} (E_n)|v^\mu|\psi_{\alpha,m} (E_m) \rangle , \nonumber
\end{align}
where $v^\mu$ is electron velocity matrix 
\begin{align}v^\mu = (ie)^{-1} \sum_{k} M^{k\mu}, \end{align}
the corresponding matrix element is 
\begin{align}\label{eq:vme2}
v^{\mu}_{mn} =\hbar^{-1} H_{mn} (r^{\mu}_m-r^{\mu}_n). 
\end{align}
In the second line of Eq.~(\ref{eq:x2}), we have written it in terms of velocity matrix elements between scattering states originated from electrode $\alpha$ and $\beta$. 
This form highlights the nature of the bubble diagram representing inelastic transitions from scattering state of one electrode $\alpha$ at higher energy $|\psi_{\alpha, m}(E)\rangle $ to that of electrode $\beta$ at lower energy $|\psi_{\beta, n}(E^-)\rangle$, with $m$ and $n$ the state index. Photon emission
is accompanied by these transitions.

We focus on angular momentum in the direction along the chain ($z$), which is the largest component, 
\begin{align}
\label{equ:JA}
&J_A(\omega) = J_z (\omega)\\
&= \Theta(eV-\hbar\omega) \frac{ \omega \alpha}{3 \pi^2 c^2} \int_{\mu_R^+}^{\mu_L} dE j_A(E,E^-), \nonumber
\end{align}
with the voltage bias $eV=\mu_L-\mu_R>0$, $\mu_R^+=\mu_R+\hbar\omega$, the fine-structure constant $\alpha = e^2/(4\pi \varepsilon_0 \hbar c) \approx 1/137$ and  
\begin{align}
j_A(E,E^-) = 2 \hbar \operatorname{Im}X_{xy}^{L R}(E,E^-).
\label{eq:jax}
\end{align}
The AMR can be analyzed through the energy dependence of $j_A(E,E^-)$. 
From Eq.~(\ref{equ:JA}) we see that the emission spectrum of the system at given energy $\hbar\omega$ can be obtained by integrating along a line cut $E$ over an effective bias window $[\mu_R+\hbar\omega, \mu_L]$ where the inelastic optical transition can take place. This bias window is determined by the relative positions between the two electrode chemical potentials, controlled by the Heaviside step function in Eq.~(\ref{equ:JA}). Thus, $j_A (E,E^{-})$ can be used to characterize the ability of the system to emit radiation with angular momentum. This is especially useful in molecular junctions where the rotational symmetry is broken and orbital angular momentum is no longer a good quantum number to characterize the symmetry property of molecular orbitals, as in simple molecules\cite{zhang_angular_2020}. 
The total AMR is then obtained by integrating the spectrum 
\begin{align}
J_A &= \int_{0}^{\infty} d \hbar\omega J_A(\omega).
\label{equ:JAi}
\end{align}
Similarly, define
\begin{align}
J_N(\omega) &= \Theta(eV-\hbar\omega) \frac{ \omega\alpha}{3 \pi^2 c^2}  \int_{\mu_R^+}^{\mu_L} {d E} j_N(E,E^-). \label{eq:JN}
\end{align}
with 
\begin{align}
j_N = {\rm Re}\left\{X^{LR}_{xx}+X^{LR}_{yy}+X^{LR}_{zz}\right\},
\end{align}
the power and number current are 
\begin{align}
P &= \int_{0}^{\infty} d\hbar\omega \hbar\omega J_N(\omega), \\
J_N &= \int_{0}^{\infty} d\hbar\omega J_N(\omega).
\label{eq:JN_int}
\end{align}

We note the difference between angular momentum current and energy/number current. They are determined by imaginary and real part of $X_{\mu\nu}^{LR}$, respectively. Moreover, for photon number or energy current we need to sum over three diagonal elements, while for the angular momentum current in $z$ direction we only need the cross term Im$X^{LR}_{xy}$. According to Eq.~(\ref{eq:x2}), we have 
\begin{align}
j_N(E,E^-) &\sim \sum_{\mu=x,y,z}|\langle \psi_{L,m}(E)|v^{\mu}|\psi_{R,n}(E^-)\rangle|^2 , \\
j_A(E,E^-) &\sim {\rm Im}\left\{\langle \psi_{L,m}(E)|v^{x}|\psi_{R,n}(E^-)\rangle\right. \nonumber\\
&\left.\langle \psi_{R,n}(E^-)|v^{y}|\psi_{L,m}(E)\rangle \right\}.
\label{eq:xyi}
\end{align}
Their form suggests that the energy and number current is directly determined by the inelastic dipole transition rates, while the angular momentum current is related to the interference between dipole matrix elements in $x$ and $y$ direction. This can be seen by writing the optical matrix element for the circular polarized light as
\begin{align}
j_{N,\pm}(E,E^-) \sim |\langle \psi_{L,m}(E)|v^{x}\pm i v^y|\psi_{R,n}(E^-)\rangle|^2 ,
\end{align}
and noticing that 
\begin{align}
j_A(E,E^-) \sim j_{N,+}(E,E^-) - j_{N,-} (E,E^-).
\end{align}
This indicates the mechanism of angular momentum radiation discussed here originates from orbital effect. Geometrical chiral properties of the molecule are encoded in the velocity matrix $v^\mu$.
It does not involve the magnetic property, distinct from the normal chiroptical response\cite{polavarapu_chiroptical_2017}. Moreover, spin and spin-orbit interaction are not prerequisite, although their inclusion may lead to new effect. 

To further understand the orbital nature of the AMR,
we notice that Eq.~(\ref{equ:JA}) is proportional to the fine structure constant $\alpha$ divided by $c^2$. When considering the non-relativistic limit of an electron undergoing equivalent spiral motion within the helical chain,
the corresponding classical angular momentum radiation can be described by the equation \cite{epp_angular_2019} 
$J_A^{\rm{c}}=2\alpha r_g^2 \hbar \omega^3_0/3c^2$.
This bears resemblance to Eq. (\ref{equ:JA}), suggesting that the AMR $J_A$ is linked to the spiral motion of electrons in both quantum and classical regime. 
We can estimate the magnetic field required that would support an equivalent spiral motion of electrons within the chain, $B \sim 5 \times 10^3 $T.  
Such high magnetic field is impossible to realize in the lab, while using the chiral molecule can avoid such huge external field.  
This illustrates the advantage of using chiral molecules for angular momentum generation. 

Since the angular momentum changes sign upon time reversal, in system with time-reversal symmetry (TRS) AMR is identically zero. Non-zero AMR needs breaking of TRS. In optical rotation and circular dichroism, the molecular eigen states are time-reversal symmetric. Breaking of TRS is realized by the external magnetic field. Meanwhile, the biased chiral molecular junction studied here is an open system. The electronic states participating the inelastic transition are scattering states. The involved scattering states are determined by current direction. The scattering states propagating in opposite directions are linked to molecular structures with opposite chirality.   
Thus, TRS breaking is realized by external bias and resulting electrical current. Magnetic field is not necessary. This is the central feature of present mechanism. It enables electrical generation of optical angular momentum utilizing the chiral geometric properties of the molecule without introducing magnetic field. It also differs from other approaches where optical angular momentum is generated by chiral wave guide from initially linear polarized light. 

Finally, we note that the form of Eq.~(\ref{eq:x2}) is similar to the expressions for nonconservative current-induced forces in molecular junctions\cite{Lu2012current,dundas2009current,wu2021electronic}. There, the flow electrical current exerts forces on atomic degrees of freedom. The nonconservative nature of the force enables angular momentum transfer to the atomic motion from electrons. Breaking of time-reversal symmetry in the electronic system by the presence of electrical current is crucial. Here, the angular momentum is transferred to photons instead of atomic motion. Thus, we envisage that the insights gained there can also be used to understand the physics of AMR studied here. 

\begin{figure}[ht]
    \centering
    \includegraphics[width=0.95\linewidth]{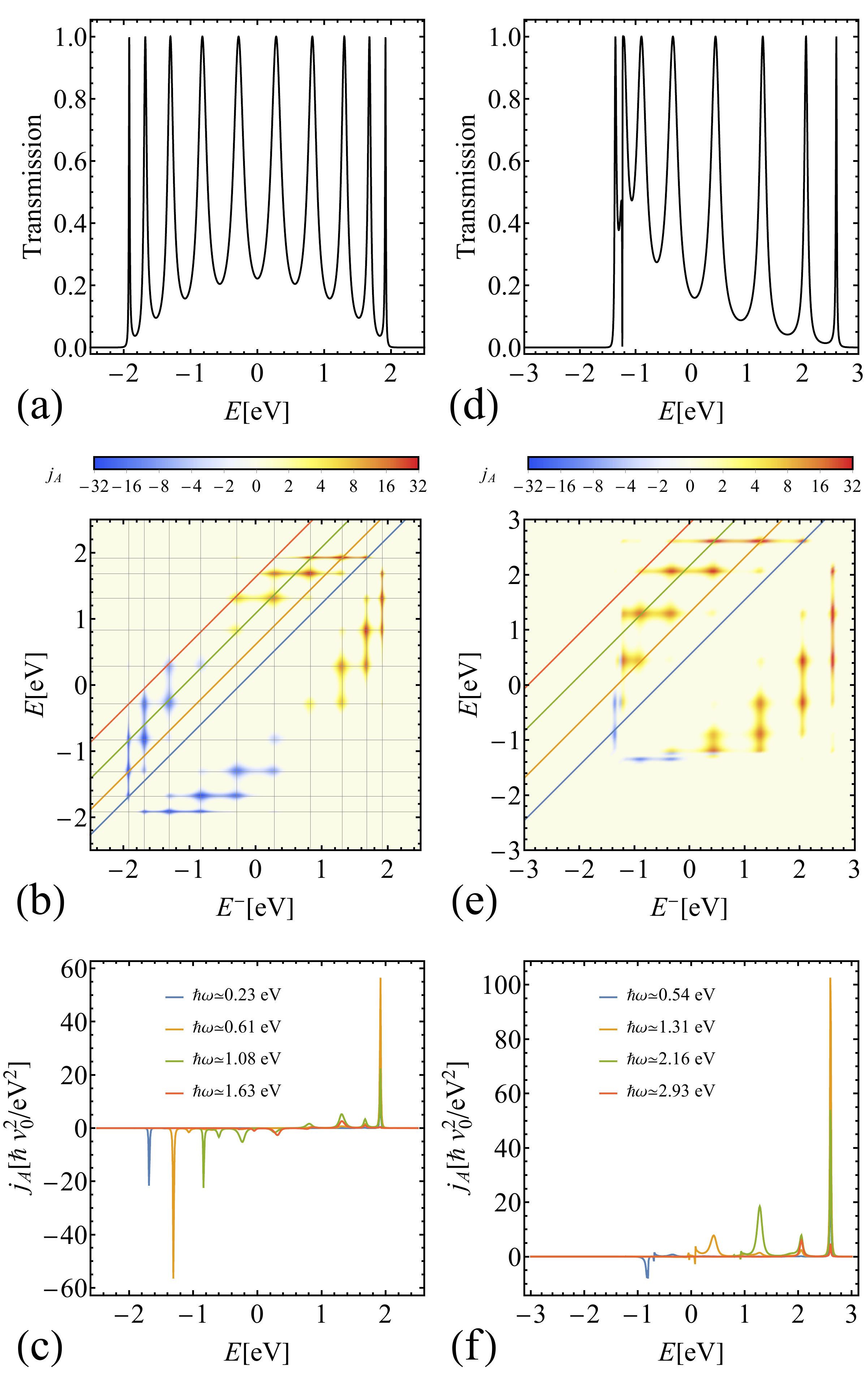}
    \caption{Similar results as Fig.~\ref{fig:mod1}, for single helical chain with $N_c=10$. The other parameters are the same as those for $N_c=3$. 
    }
    \label{fig:s10_nn_nnn}
\end{figure}

\section{\label{sec:results}Numerical results}

We now present numerical results for model helical chain structures. We consider zero temperature case $T=0$. The default geometric parameters are radius of gyration $r_g = 7 \textup{~\AA} $, arc length $l_a =5.6\textup{~\AA}$,
helix angle $\theta \approx 0.66$, and phase angle $\Delta\phi = \pi/5$. 
The electronic structure is modeled by using tight-binding parameters, with onsite energy set to zero, the nearest neighbour (NN) hopping $t_{\rm NN}=1$ eV, different values of the next nearest neighbour (NNN) hopping $t_{\rm NNN}$ will be used. 
An energy step of $0.1$ meV is used to do the numerical integration, and the energy range is set to $[-5, 5]$ eV. 
To consider coupling of the chain to electrodes, 
we use the wide band limit here with an energy independent retarded self-energy for each electrode
\begin{align}
    \Sigma^{r/a}_{L/R} &= \mp i \Gamma_{L/R}/2 .
\end{align}
The broadening matrix $\Gamma_{L/R}$ is diagonal. The $m$-th diagonal matrix element $\Gamma_{L/R, mm}$ takes the same constant non-zero value $\gamma=0.1$ eV if the site $i$ couples to the electrode and zero otherwise.

\subsection{Angular momentum radiation spectrum}
\label{para:Term X for default configuration}
Note that to analyze the full radiation spectrum we consider the large bias limit in this subsection, unless specified. This means that the chemical potential is set to $\mu_L > {\rm max}\{E_m\} > {\rm min}\{E_m\} > \mu_R$ so that all possible inelastic transitions can be activated. Here $\{E_m\}$ are the set of energies of molecular eigen state given by diagonalizing $H_0$, and coupling to electrodes leads to broadening of the states.

We start from the simplest structure of a single helical chain with length
$N_c=3$. Figure~\ref{fig:mod1} summarizes the main results without (a-d) and with
(e-h) NNN hopping, respectively. The central molecule with $N_c=3$ has three
energy eigen states. The three states are symmetric with respect to zero in the
NN case. When the equilibrium chemical potential is zero, the system has
electron-hole symmetry. The optical transition $E_3 \to E_2$ and $E_2
\to E_1$ emit photons with the same energy but opposite angular momentum. Thus,
they cancel with each other. This symmetry is broken after including NNN
hopping. This is reflected in the electron transmission spectra shown in
Fig.~\ref{fig:mod1} (a, e) and the schematic diagram in (d, h). The cancellation
among different transitions is not perfect anymore, leading to larger AMR.
Figure~\ref{fig:mod1} (b, f) show the corresponding AMR spectra $j_A (E, E^-)$.
The corresponding line cuts for given photon energy $\hbar\omega$ are shown in
(c, g). 
The sharp peaks that dominate the AMR contribution in the parameter space ($E$, $E^-$) correspond to inelastic transition between different molecular eigen states. The positive and negative values of $j_A$ correspond to opposite AMR, which is schematically shown in Fig.~\ref{fig:mod1}(d,h) with different colors.
The situation is similar when the bias is reversed, but the AMR also changes sign, i.e., the positive and negative values are switched.  

    \begin{figure}[ht]
    \centering
    \includegraphics[width=0.5\textwidth]{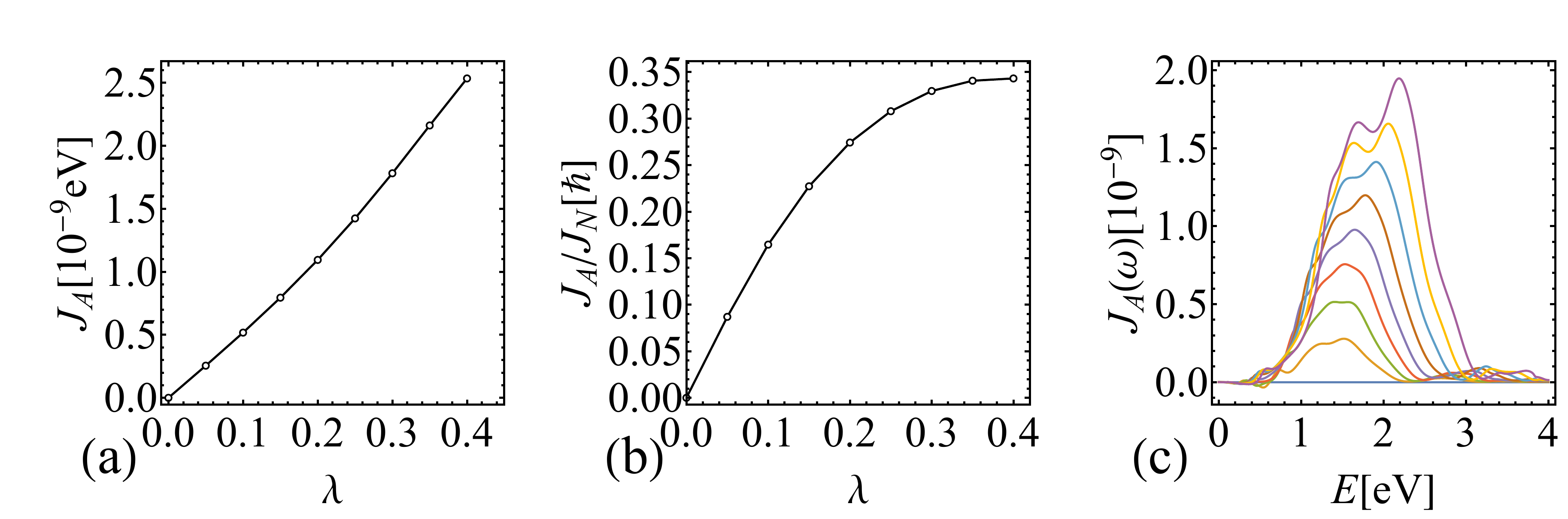}
    \caption{ AMR ($J_A$) (a), AMR per photon ($J_A/J_N$) (b) for single helical chain with $N_c=10$ as a function of $\lambda=t_{\rm NNN}/t_{\rm NN}$ in the large bias limit $\mu_L > {\rm max}\{E_m\} > {\rm min}\{E_m\} > \mu_R$, with $\{E_m\}$ the set of eigen energies of the molecular orbitals. 
     (c) AMR spectrum $J_A(\omega)$ from $\lambda=0$ (bottom) to $0.4$ (top). }
    \label{fig:ssdna_neighbor}
    \end{figure}

Results for a more realistic chain with $N_c=10$ are shown in Fig.~\ref{fig:s10_nn_nnn}.
The AMR distribution represented by $j_A(E, E^-)$ spreads to much larger regions.
The electron-hole symmetry breaking due to NNN hopping is more obvious than the shorter chain. The hole contribution (negative $E$) decreases and the whole distribution is dominated by the electron contributions (positive $E$). This is also reflected in the asymmetric distribution of the electron transmission in the positive and negative energy range.  We have shown in Fig.~\ref{fig:ssdna_neighbor} the dependence of $J_A$ on the NNN hopping in the large bias limit. We observe increase of both magnitude and efficiency of AMR with the relative magnitude of NNN hopping $\lambda=t_{\rm NNN}/t_{\rm NN}$, and the efficiency $J_A/J_N$ saturates at around $0.3 \hbar$.

\begin{figure}[ht]
    \centering
    \includegraphics[width=1.0\linewidth]{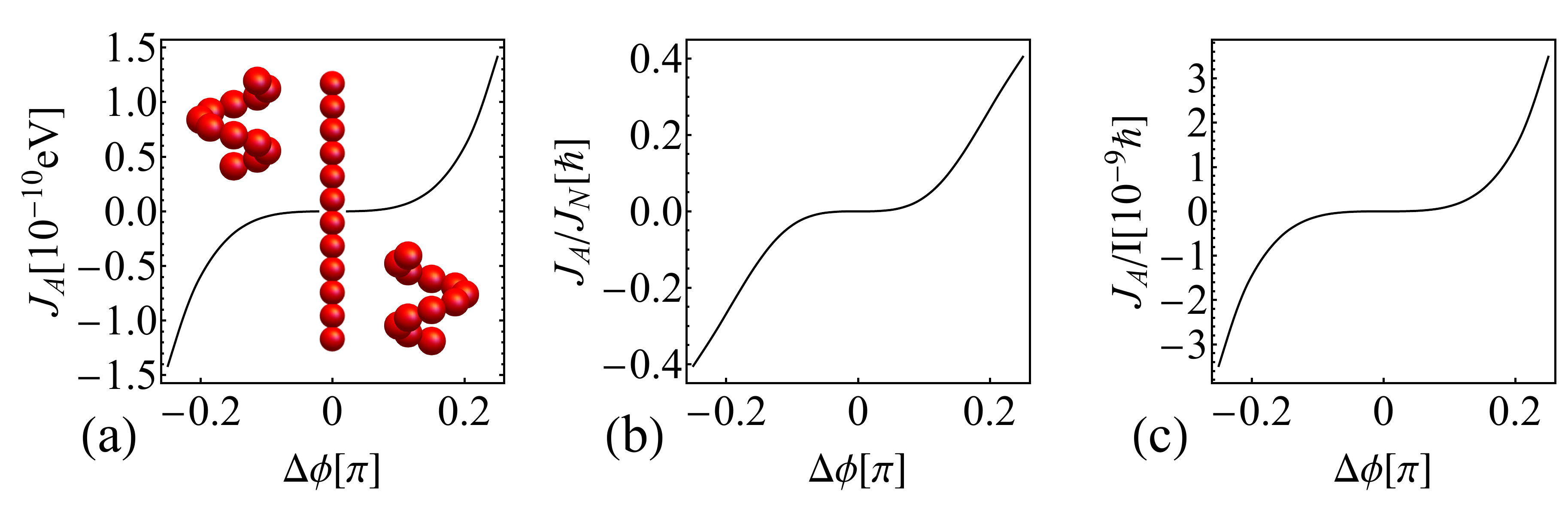}
    \caption{ 
        Dependence of total AMR $J_A$ (a), AMR per photon $J_A/J_N$ (b) , AMR per electron $J_A/I$ (c) as a function of phase angle $\Delta\phi$ for parameters $N_c=12, \lambda=0.2, \mu_L=-\mu_R=2$ eV. Insets of (a) show single helical chains with phase angle $\Delta\phi = -\pi/4, 0, \pi/4$, respectively. Comparing (b) and (c) shows that, although  the photon emission efficiency is low, the average angular momentum carried by each photon can be quite high. 
    }
    \label{fig:ssdna_phase_angle}
\end{figure}

\begin{figure}[ht]
    \centering
    \includegraphics[width=0.5\textwidth]{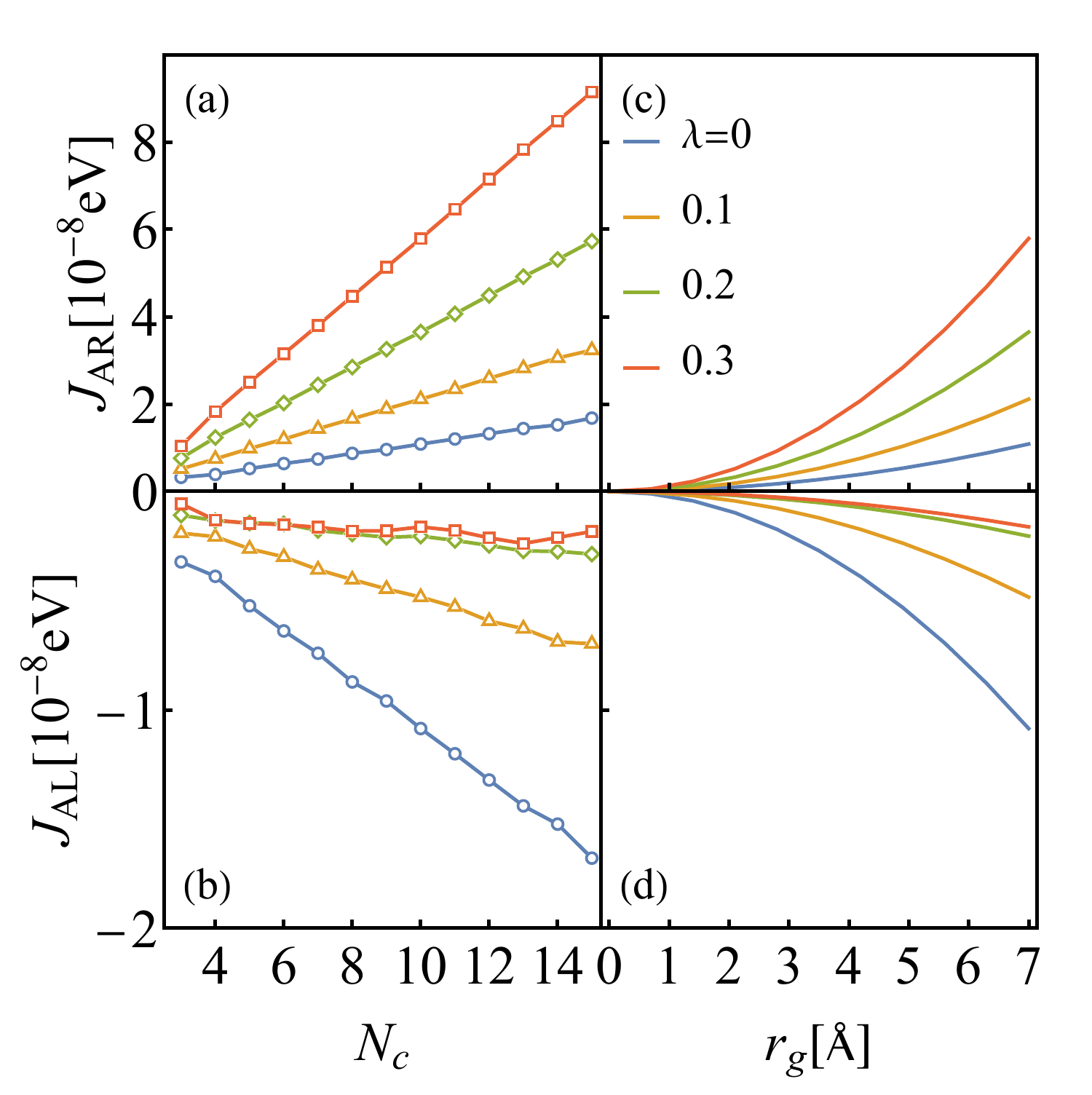}    
    \caption{ 
        AMR as a function of chain length $N_c$ (a-b) and radius of gyration $r_g$ (c-d) with different $\lambda$ in the large bias limit. Default values are used for other parameters. 
        $J_{\rm AR}$: RH radiation corresponding to $j_A>0$,  $J_{\rm AL}$: LH radiation  corresponding to $j_A<0$. Note the different scales of $J_{\rm AR}$ and $J_{\rm AL}$.
    }
    \label{fig:s10_geo_N_R}
\end{figure}

\subsection{Geometrical dependence}
\label{para:Geometry Chirality}

To show the geometrical origin of the AMR in chiral molecules, 
dependence of AMR in the high bias limit on phase angle $\Delta\phi$ is shown in Fig.~\ref{fig:ssdna_phase_angle}. It can be seen that AMR increases with the absolute value of phase angle. It is exactly zero for achiral straight chain and changes sign when $\Delta \phi$ goes from positive to negative. 
Further analysis of $j_A (E, E^-)$ shows that the whole distribution in energy space reverses sign when $\Delta\phi$ changes sign. 
This positive correlation between AMR and chirality of the chain is the first evidence of geometrical origin of AMR studied here. It can be understood from the velocity matrix element [Eq.~(\ref{eq:vme2})]. When the phase angle changes, the relative position of neighbouring atoms also changes. The chiral information is encoded in the relative positions. In the case of straight chain,  $j_A$ can be shown to be exactly zero, and 
it changes sign when the phase angle passes zero.

We depict the length ($N_c$) and radius ($r_g$) dependence of the AMR in Fig.~\ref{fig:s10_geo_N_R}. We have integrated all the positive (termed right-handed (RH), $j_{\rm AR}$) and negative (termed left-handed (LH), $j_{\rm AL}$) regions of the $j_A$ spectra in the energy space of $(E, E^-)$ to characterize the system's ability to radiate angular momentum. 
The AMR grows linearly with chain length and quadratically with radius. This can be explained by velocity matrix $v^\mu$, which distributes along the whole chain. As the chain length increases, the velocity matrix element between the scattering states also increases. This leads to a linear dependence on the chain length. As for radius, $v^\mu$ is proportional to $r_g$ and there are two $v^\mu$ factors in Eq.(\ref{eq:x2}). 
This is the second evidence of geometrical nature of AMR. 

\begin{figure}[ht]
    \centering
    \includegraphics[width=1.0\linewidth]{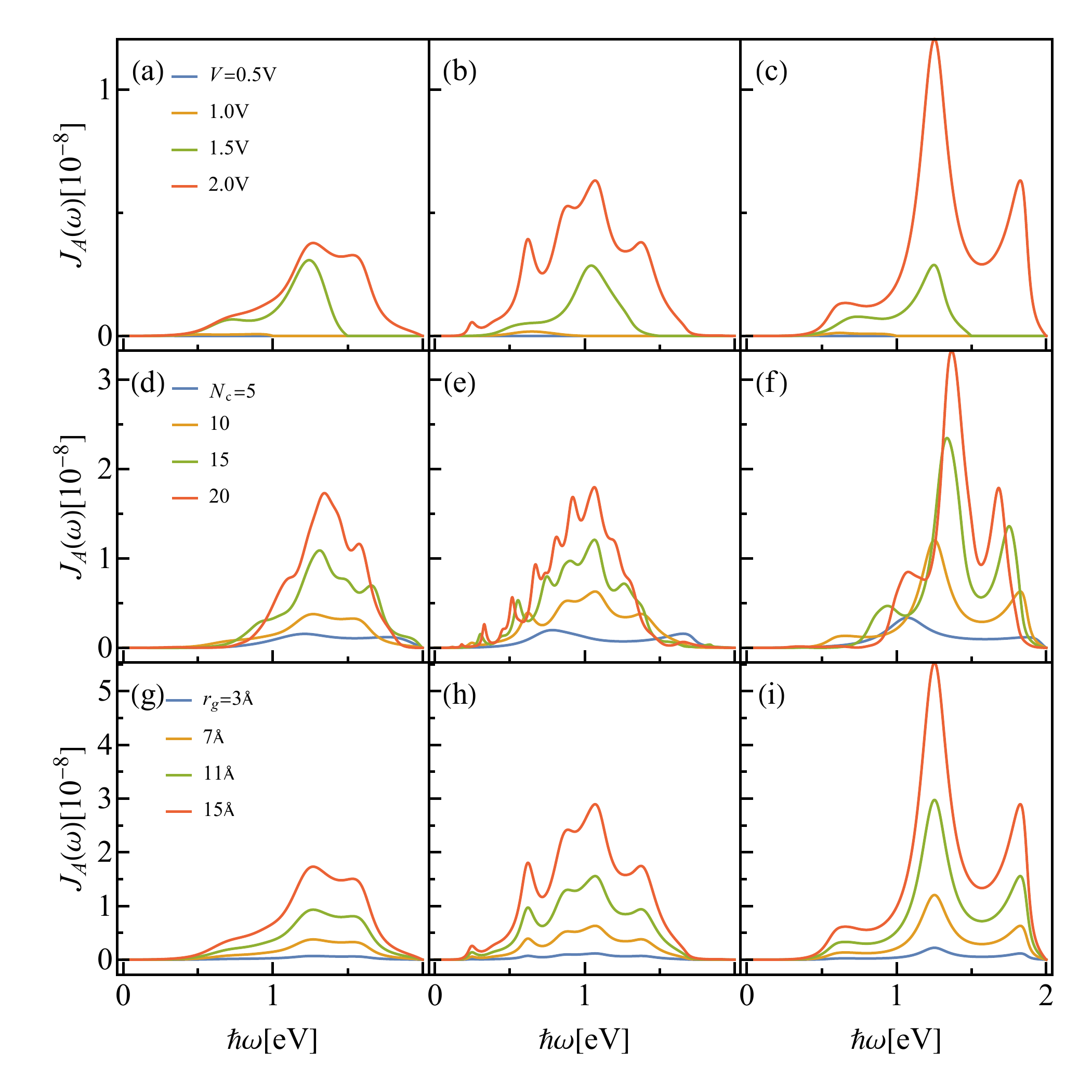}
    \caption{ (a-c) AMR spectra at different applied bias for chain $N_c=10$. (d-i) AMR spectra for different chain length $N_c=5, 10, 15, 20$ (d-f) and radius of gyration $r_g = 3, 7, 11, 15$~\AA~(g-i) at $V=2$ V. The three columns from left to right correspond to results under symmetric bias ($\mu_L=-\mu_R=|eV|/2$) with NNN hopping, asymmetric bias ($\mu_L=|eV|, \mu_R=0$) without and with NNN hopping ($\lambda=0.2$), respectively.  }
    \label{fig:bias}
\end{figure}

\subsection{Bias dependence of angular momentum radiation}
Up to now, we have been focused on the large bias limit where all the inelastic transitions contribute to AMR. We now consider more realistic cases at different biases for two types of voltage drop with and without NNN hopping. The results are depicted in Fig.~\ref{fig:bias}. For symmetric drop ($\mu_L=-\mu_R=|eV|/2$) with NN hopping, the AMR is zero due to electron-hole symmetry of the system. Thus, it is not shown here. Inclusion of NNN hopping breaks this symmetry and non-zero AMR is observed (Fig.~\ref{fig:bias}a, d, g). The spectra width grows with the applied bias and the high energy cutoff is given by $|eV|$ (Fig.~\ref{fig:bias}a). When an asymmetric bias is applied, with $\mu_L=|eV|, \mu_R=0$, AMR can be obtained both without (Fig.~\ref{fig:bias}b, e, h) and with (Fig.~\ref{fig:bias}c,f,i) NNN hopping. The magnitude of AMR becomes larger than the corresponding symmetric bias cases, especially at $V=2.0$ V. This is understandable from the 2D plot of $j_A$ in Fig.~\ref{fig:s10_nn_nnn} (b, e). The range of the energy integral $\mu_R+\hbar\omega \le E \le \mu_L$ in the asymmetric case only covers the upper panel, which is positive. Meanwhile, in the case of symmetric voltage drop, the integral includes both positive and negative contributions and they cancel with each other. Introducing the NNN hopping further makes the positive contribution larger and leads to larger AMR. We have also studied the length (d, e, f) and radius  (g, h, i) dependence of AMR at different biases. The results are consistent with those presented in Fig.~\ref{fig:s10_geo_N_R}, i.e., AMR grows with chain length and radius of the helical chain.

\section{\label{sec:conclusion}Conclusion}
In summary, we have studied electrically driven angular momentum radiation (AMR) from helical chains using the nonequilibrium Green's function method. 
The ability of AMR is characterized by the imaginary part of a joint optical transition matrix element between scattering states originated from the two electrodes [Eq.~(\ref{eq:x2})].  
We have made direct connection between the geometrical factors and the radiation properties. 
The most important property of this chiral-induced AMR is that it does not rely on the magnetic field. Rather, it relies on the electrical dipole transitions at two different directions, from filled to empty scattering states originated from two different electrodes. We have also shown the dependence of AMR on the tight-binding parameters and the coupling to electrodes. These parameters allow electrical engineering of molecule's AMR property. 
This fully electrical way of generating AMR employing chiral molecules may find its usefulness in the development of chiral single molecule light sources.

\begin{acknowledgments}
We thank Prof. Jian-Sheng Wang (NUS) for valuable discussions. This work was supported by National Natural Science Foundation of China under Grants No. 22273029 and No. 21873033.
\end{acknowledgments}

\bibliographystyle{apsrev4-2}
\bibliography{main2024}

\end{document}